# BCI-sift: An automated feature selection toolbox for Brain-Computer Interface applications

Elena C Offenberg[1†], Dirk Keller[1†], Mariska J Vansteensel[1], Zachary V Freudenburg[1], Nick F Ramsey[1,2] and Julia Berezutskaya[1]

[†] Equal Contribution
[1] Department of Neurology and Neurosurgery, University Medical Center Utrecht Brain Center, Utrecht University, Utrecht, The Netherlands
[2] Donders Institute for Brain, Cognition and Behaviour, Radboud University, Nijmegen, The Netherlands

Corresponding Author: Julia Berezutskaya
E-mail: y.berezutskaya-2@umcutrecht.nl

**Abstract**

*Objective*
Advancements in Brain-Computer Interfaces (BCIs) for clinical applications rely on precise and reliable signal interpretation. However, the high-dimensional and noisy nature of data captured from both implanted and non-implanted BCIs presents significant challenges, necessitating the use of sophisticated filtering techniques, such as feature selection algorithms. We introduce the Python-based BCI-sift (BCI Systematic and Interpretable Feature Tuning) Toolbox, a comprehensive tool designed to streamline the application of diverse optimization algorithms to BCI datasets for identifying the most relevant features in machine learning tasks.
*Approach*
Our scikit-learn-compatible toolbox (github.com/UMCU-RIBS/BCI-sift) simplifies feature selection in BCI tasks by integrating advanced optimization methods. It was validated on data from eight able-bodied participants implanted with 64 to 128 high-density electrocorticography (HD ECoG) electrodes. The electrodes were placed on the sensorimotor cortex (SMC), and participants repeatedly spoke 12 different words.
*Main Results*
BCI-sift successfully identified informative neural features across time points, electrodes, and frequency dimensions. The anatomical locations of electrode selections were consistent across participants and aligned with the functional and anatomical organization of motor activity in the SMC. The most relevant time points were determined to be during word pronunciations. In the frequency dimension, BCI-sift identified the high frequency band as most informative for classification, consistent with prior research. Beyond providing interpretability, BCI-sift significantly improved classification accuracy compared to using no feature selection in all dimensions.
*Significance*
With BCI-sift, we provide an accessible and versatile platform for feature selection in BCI research, enabling improved classification accuracy, automatic feature analysis, and enhanced interpretability. Although validated with HD ECoG data, the underlying principles are applicable to other data, including implanted and non-implanted BCIs. We encourage researchers to explore and apply BCI-sift in these contexts as well. By enhancing classification accuracy and interpretability, BCI-sift addresses key challenges in developing efficient and transparent BCI systems.

**Acknowledgements**

This publication is part of the project Dutch Brain Interface Initiative (DBI2) with project number 024.005.022 of the research programme Gravitation, which is financed by the Dutch Ministry of Education, Culture, and Science (OCW) via the Dutch Research Council (NWO). In addition, this project is funded by the European Union's HORIZON-EIC-2021-PATHFINDER CHALLENGES program under grant agreement No 101070939 and by the Swiss State Secretariat for Education, Research and Innovation (SERI) under contract number 22.00198.

MJV was a consultant for GA Capital. There are no further conflicts of interest for any of the authors.

## 1. Introduction

Accurate and robust signal interpretation is central to both deepening our understanding of brain functions and driving progress in the development of brain-computer interfaces (BCIs). Recent advances in both implanted and non-implanted recording techniques, such as high-density (HD) and ultra-high-density electrocorticography (ECoG) grids (Hettick et al., 2022), microelectrode arrays (MEAs) (Rubin et al., 2023), and high-density EEG (Stoyell et al., 2021), have enabled the collection of increasingly large and complex neural datasets. While this increase in data volume holds the promise of improved decoding performance, it also introduces substantial analytical challenges. In particular, the extraction of relevant information from high-dimensional signals requires advanced preprocessing methods, including feature selection algorithms. These techniques are crucial for enhancing decoding accuracy, reducing computational demands, and improving the interpretability of BCI systems.

Feature selection serves several critical functions in BCI systems. As the volume and dimensionality of neural data increase, particularly with high-resolution recording technologies, many extracted features may be redundant or unrelated to the mental states of interest (Altan et al., 2021, Wang et al., 2025). Reducing the number of input features can alleviate this issue, enabling classifiers to operate on a more relevant subset of data. This not only enhances computational efficiency, but also decreases the number of parameters that need to be estimated, thereby mitigating the risk of overfitting, especially when training data is limited (Defernez & Kemsley, 1999). As a result, BCI training for new users may require less data, enabling faster and more reliable system deployment. Moreover, selecting a compact set of informative features, or assigning relative importance to features, can facilitate a clearer association between neural activity and cognitive or motor states, ultimately improving the interpretability of BCI models.

The identification of optimal features has been a central topic of research, particularly in the domain of EEG-based BCI systems, where channel selection forms an integral part of the signal processing pipeline (Arvaneh et al., 2011). Techniques such as recursive feature elimination, along with more advanced biologically inspired methods including genetic algorithms and particle swarm optimization, have been employed to identify the most informative electrodes (Wei et al., 2006, Wei & Wei Tu, 2008, Offenberg et al., 2025). Although some studies have also investigated feature relevance in specific frequency bands (Kabir et al., 2024), this aspect has received comparatively less attention. In implanted BCI systems, deep learning approaches have been used in feature optimization for performance gains (Berezutskaya et al., 2023, Hosman et al., 2023, Karpowicz et al., 2025), and optimization methods such as genetic algorithms have also been adapted to this setting (Idowu et al., 2020). Despite these advances, the overall landscape remains fragmented, with most approaches developed for a single recording modality or constrained to specific application contexts. In addition, many methods are evaluated in only a single participant, limiting their potential for broad clinical translation of BCIs. To the best of our knowledge, until now, there has not been a single, easy-to-use tool that can perform feature selection for a wide range of contexts.

We introduce BCI-sift (BCI Systematic and Interpretable Feature Tuning), a Python-based toolbox designed to streamline feature selection in BCI tasks by incorporating a range of advanced optimization techniques. The toolbox includes dedicated algorithms for contiguous feature selection, namely an exhaustive search and a stochastic hill climbing approach, which operate on continuous windows in one dimension or rectangular subregions in two-dimensional feature layouts. In addition, BCI-sift integrates several general optimization methods, including recursive feature elimination, simulated annealing, evolutionary strategies, and particle swarm optimization. The toolbox is intended to offer researchers an accessible and flexible framework for exploring BCI-related datasets and identifying the most informative features across various dimensions, including electrode, time, and signal frequency. BCI-sift is fully compatible with the scikit-learn ecosystem (Pedregosa et al., 2018), allowing it to be integrated with a wide range of prediction algorithms. Its modular architecture facilitates easy extension to additional optimization methods and is intended to support ongoing development and contributions from the broader research community. The toolbox is openly available at https://github.com/UMCU-RIBS/BCI-sift, together with documentation and a tutorial.

As validation, we applied the toolbox to a classification task that involved eight able-bodied participants pronouncing 12 Dutch words 10 times each, while neural data was recorded using HD ECoG grids. Based on our validation analysis, we found that our toolbox provided the following potential benefits: increased decoding accuracy, identification of relevant features in the frequency, electrode and time dimensions, overcoming task constraints, and improved interpretability of the dataset.

## 2. Methods

### *2.1 Feature selection procedure*

Within the BCI-sift workflow, brain data from a BCI task, the corresponding labels and analysis parameters (as defined by the user) are provided as inputs to the toolbox (Figure 1). The analysis parameters specify the feature dimension(s) to optimize, the optimization algorithm(s) to apply, the scikit-learn pipeline containing the analysis model, the evaluation metric, and any algorithm-specific hyperparameters. During optimization, the toolbox repeatedly trains the model using the subset of features selected at each step. For every selection, it records both the corresponding feature mask and the resulting evaluation metric. Depending on user preference, specified in the algorithm parameters, the evaluation metric can be cross-validated or not. Once optimization is concluded, the toolbox returns the feature mask associated with the highest performance, which can then be applied to new data. Optionally, users can add an outer cross-validation loop, enabling the selected masks to be tested on held-out data within the same dataset for a more robust evaluation.

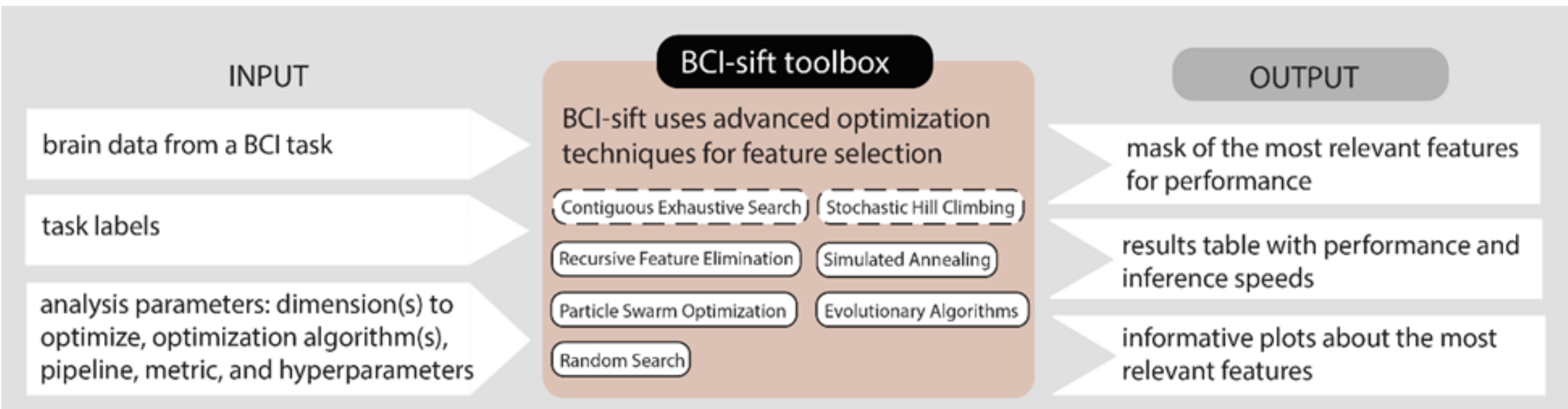

**Figure 1. Overview of the BCI-sift workflow.** The toolbox receives brain data from a BCI task, along with corresponding labels and user-defined analysis parameters for the feature selection procedure. BCI-sift then performs the selected optimization analysis and provides the following outputs: a feature mask indicating the most relevant features, a results table summarizing the evaluation metrics, and plots for result interpretation. Contiguous optimization techniques are indicated by dashed outlines, whereas general optimization methods are shown with solid outlines.

### *2.2 Contiguous search algorithms*

In many BCI recording modalities, such as microelectrode arrays, EEG or ECoG, data is recorded from electrodes arranged in a two-dimensional layout, creating an inherent spatial structure in the data. Exploiting this structure during feature selection can provide insights that are directly relevant for electrode placement and device design. In particular, identifying spatially contiguous regions that contribute most to task performance can inform which electrode subregions are most informative. Similarly, in other feature dimensions such as time, it can be useful to determine which continuous segments, for example a temporal window around an event, carry the strongest discriminative information. To support this type of analysis, the contiguous optimization methods in our toolbox restrict the search space to contiguous feature subsets. Instead of evaluating arbitrary combinations of features, the algorithms test only continuous windows in one dimension (e.g., time) or rectangular subselections in two dimensions (e.g., smaller electrode grids within the original electrode layout). This enables the identification of compact, interpretable regions of informative features and can provide guidance for the planning of future electrode placements.

The *Contiguous Exhaustive Search* method systematically evaluates all possible contiguous feature segments, selecting the configuration that yields the highest cross-validated performance. In one-dimensional cases, this corresponds to testing all possible continuous windows (e.g., time segments), while in two-dimensional cases it evaluates all possible rectangular subgrids of electrodes. As a computationally more efficient alternative, *Stochastic Hill Climbing* iteratively explores this contiguous search space. The algorithm begins with a minimal configuration, namely a single feature in one dimension or a 1x1 electrode grid in two dimensions, and progressively expands the selected region to adjacent features. At each iteration, the expansion direction is chosen using one of two strategies: exploration, where a random neighboring direction is selected, or exploitation, where all candidate directions are evaluated and the one producing the largest improvement in the objective function is chosen. The balance between exploration and exploitation is controlled by an exploration parameter (ε), which gradually shifts the algorithm's behavior from random exploration toward more directed exploitation as the search progresses. The procedure continues until the search space has been traversed, and the configuration that achieved the highest value of the objective function is returned as the final solution.

### *2.3 General feature selection algorithms*

The remaining algorithms in the toolbox are applicable to any feature dimension and do not require selected features to be spatially or temporally adjacent. These methods differ primarily in how they navigate the often large and complex space of possible feature combinations. In the case of *Recursive Feature Elimination* (RFE), the user provides an importance metric that assigns weights to individual features based on the trained model. One example of such an importance metric is using the coefficients of a support vector machine (SVM) with a linear kernel. RFE proceeds by iteratively removing the feature with the lowest absolute weight, assuming it contributes least to classification performance. Although RFE offers an intuitive and computationally efficient approach, it is limited to models that produce interpretable feature weights. As a result, it is not compatible with more complex models, such as non-linear neural networks, which do not provide straightforward measures of feature importance.

The other general feature selection algorithms are compatible with any prediction model, as they optimize feature selection based solely on the user-defined evaluation metric. Each algorithm differs in how it explores the space of possible feature combinations, using a method-specific search strategy. *Evolutionary Algorithms* iteratively refine a population of candidate solutions using operations such as mutation, crossover, and selection. In this context, mutation introduces random changes to a subset of features to maintain diversity in the search space, crossover combines elements from two high-performing feature subsets to generate new candidates, and selection retains the best-performing subsets based on their objective function scores. These algorithms are robust and adaptable across a wide range of optimization tasks, but can be computationally demanding due to their population-based nature (Bäck, 1996). The implementation of evolutionary algorithms in BCI-sift utilizes the DEAP-library (Fortin et al., 2012). *Simulated Annealing* is inspired by the physical annealing process in metallurgy (Kirkpatrick et al., 1983). The algorithm begins with an initial feature configuration and iteratively proposes new configurations by introducing small random modifications, such as adding or removing features. Acceptance of a worse configuration is governed by a temperature-dependent probability that decreases over time according to a cooling schedule. As the temperature lowers, the algorithm transitions from broad exploration to focused exploitation, thereby improving convergence toward a near-optimal feature subset. *Particle Swarm Optimization*, inspired by the collective behavior of bird flocks and fish schools, optimizes feature selection by coordinating a population of candidate solutions (particles) (Kennedy & Eberhart, 1995). Each particle adjusts its position in the search space based on its own best performance and that of the swarm's global best. These updates balance exploration and exploitation through weighted influences of individual and collective experience, allowing the swarm to converge toward feature subsets that maximize the objective function. Particle swarm optimization typically requires fewer parameters than evolutionary algorithms and often converges more quickly than simulated annealing. However, it may suffer from premature convergence if diversity within the swarm is lost. In BCI-sift, it is implemented with the PySwarms library (James V. Miranda, 2018). Lastly, *Random Search* selects feature subsets at random, sampling the high-dimensional search space uniformly. This approach serves as a baseline method, allowing users to benchmark the performance of other optimization algorithms within the toolbox.

### *2.4 Data collection and preprocessing*

Participants P1-P7 had medication-resistant epilepsy and were implanted with ECoG grids for clinical diagnostic purposes. During the same surgery, they were implanted with HD ECoG grids for research, and tasks were performed while participants were being monitored in preparation for epilepsy resection surgery. Participant P8 had a HD ECoG grid placed temporarily on the SMC during awake tumor resection surgery, with ECoG data being recorded intraoperatively. All participants completed a word repetition task in which isolated Dutch words were presented individually on a screen. Participants were instructed to read each word aloud, with a total of twelve unique words repeated ten times each. The dataset used for validating the toolbox has been described previously; full details regarding data acquisition, electrode localization, task design, and preprocessing are available in (Offenberg et al., 2025). The research was approved by the Medical Ethics Committee of the University Medical

Center Utrecht and conducted in accordance with the Declaration of Helsinki (2024). All participants provided written informed consent prior to participation.

One key difference in preprocessing compared with earlier analyses (Offenberg et al., 2025) lies in the normalization procedure. In previous work, neural recordings were normalized using two seconds of rest at the beginning of each run as a baseline for z-scoring. In the present study, normalization was instead implemented within the scikit-learn pipeline, ensuring that it was automatically applied at each step of the optimization process. For this purpose, a MinMaxScaler was employed. Furthermore, for the results reported in Section 3.3, common average referencing was incorporated into the pipeline, such that rereferencing was performed dynamically based on the subset of electrodes selected during optimization.

For the frequency band optimization, neural signals were decomposed into five frequency bands: delta (0.5–3 Hz), theta (4–7 Hz), alpha (8–12 Hz), beta (13–30 Hz), and the high frequency band (HFB, 70–170 Hz). For sections 3.1 and 3.2, electrodes that had not been excluded manually due to noisy or flat signals were included, with the number of available electrodes ranging from 60 in participant P6 to 128 in participant P1. A fixed time window from 1 s before to 1 s after voice onset was used for classification.

### *2.5 Pipeline Configuration*

All algorithms in the toolbox operate on a scikit-learn estimator or pipeline, which specifies both optional preprocessing steps and the prediction model. In our experiments, the pipeline included a MinMaxScaler and, for Section 3.3, common average referencing. For classification, we used a support vector machine (SVM) classifier with a linear kernel. We set the regularization parameter C to the value of $10^5$, based on previous empirical results (Offenberg et al., 2025).

For reasons of brevity, we report results mostly from the RFE algorithm, as it achieved the best performance on the dataset under study. It should be noted, however, that outcomes may vary across datasets. Moreover, RFE is limited to classifiers that assign weights to individual features, which excludes more complex models such as SVMs with non-linear kernels and non-linear neural networks. The toolbox is therefore designed to support a much broader range of optimization strategies than illustrated here, and the results presented should be regarded as an illustrative example.

### *2.6 Visualization of electrode selection results*

To visualize electrode selection results on a common cortical surface, all electrode locations were projected to Montreal Neurological Institute (MNI) space using participant-specific affine transformation matrices obtained with SPM8 (Wellcome Centre for Human Neuroimaging, University College London). Electrodes from all participants were then displayed on a standard brain surface in MNI space.

To illustrate the spatial distribution of selected electrodes, each electrode location was represented using a two-dimensional Gaussian kernel centered at the electrode coordinate (FWHM = 10 mm). This smoothing step allows nearby electrodes across participants to contribute to a continuous spatial representation, making it easier to visualize consistent regions of electrode selection across the cortex.

## 3. Results

### *3.1 Enhancing decoding performance and interpretability*

In this section, we evaluated how applying the proposed optimization algorithms across the frequency, electrode, and time dimensions improved decoding accuracy and whether it revealed the most informative neural features, with a primary focus on recursive feature elimination. Selection features across all three dimensions simultaneously would result in a very large feature space (electrodes × frequency bands × time points) and substantially increase the computational cost. Therefore, we adopted a staged optimization strategy. As an initial analysis, we applied RFE in the combined electrode-frequency domain to identify the spectral features that contribute most strongly to decoding performance. Subsequent analyses then focused on the electrode and temporal dimensions separately.

In the first elimination analysis, each electrode-frequency combination was treated as a single feature. The feature selection procedure led to a significant improvement in classification accuracy compared to classification on all electrodes and frequencies (from 19±4% to 67±19%, $Z_{all\text{-}selected}$= -2.5, p = 0.01, Wilcoxon signed-rank test, Figure 2, Supplementary Table 1 Columns 1 & 2). Figure 3 shows feature elimination patterns across frequency bands for individual participants, where electrode-frequency features are aggregated by frequency band. Across participants, the HFB was consistently the last frequency band eliminated during the recursive feature elimination process, indicating its dominant role in word classification. In participant P3, the HFB and beta band were removed concurrently, whereas in all other participants the beta band was ranked second in importance after the HFB. In contrast, the alpha, theta, and delta bands were eliminated early in the optimization, suggesting a lower contribution to classification performance. Pooled elimination sequences across participants confirmed the consistent pattern of delta being removed first, followed by theta, alpha, beta, and finally HFB (Figure 4).

In the subsequent analysis, which focused exclusively on the electrode dimension, only HFB features were included. At each optimization step, a single electrode was removed. As in the previous analysis, classification accuracy improved significantly after feature selection, with models using the selected subset of HFB electrodes outperforming models that used all electrodes in the HFB (from 66±13% to 75±19%, $Z_{all\text{-}selected}$= -2.5, p = 0.01, Wilcoxon signed-rank test, Figure 5, Supplementary Table 1 Columns 3 & 4). Across participant-specific results (Figure 6), a consistent pattern emerged wherein electrodes located closest to the central sulcus were most frequently retained in the final feature selections, indicating their importance for word classification. To visualize this pattern across participants, the electrode selection ratio, defined as the proportion of cross-validation folds in which a given electrode was included in the final feature selection, was projected onto a common MNI brain (Figure 7). To quantify the spatial trend, we computed a Spearman correlation between the electrode selection ratio and the distance to the central sulcus (approximated as $y = -20$ in MNI space). This analysis revealed a negative correlation ($\rho = -0.12$, $p = 0.0002$), indicating that electrodes closer to the central sulcus were more frequently selected. These results suggest that electrodes adjacent to the central sulcus contributed most strongly to decoding.

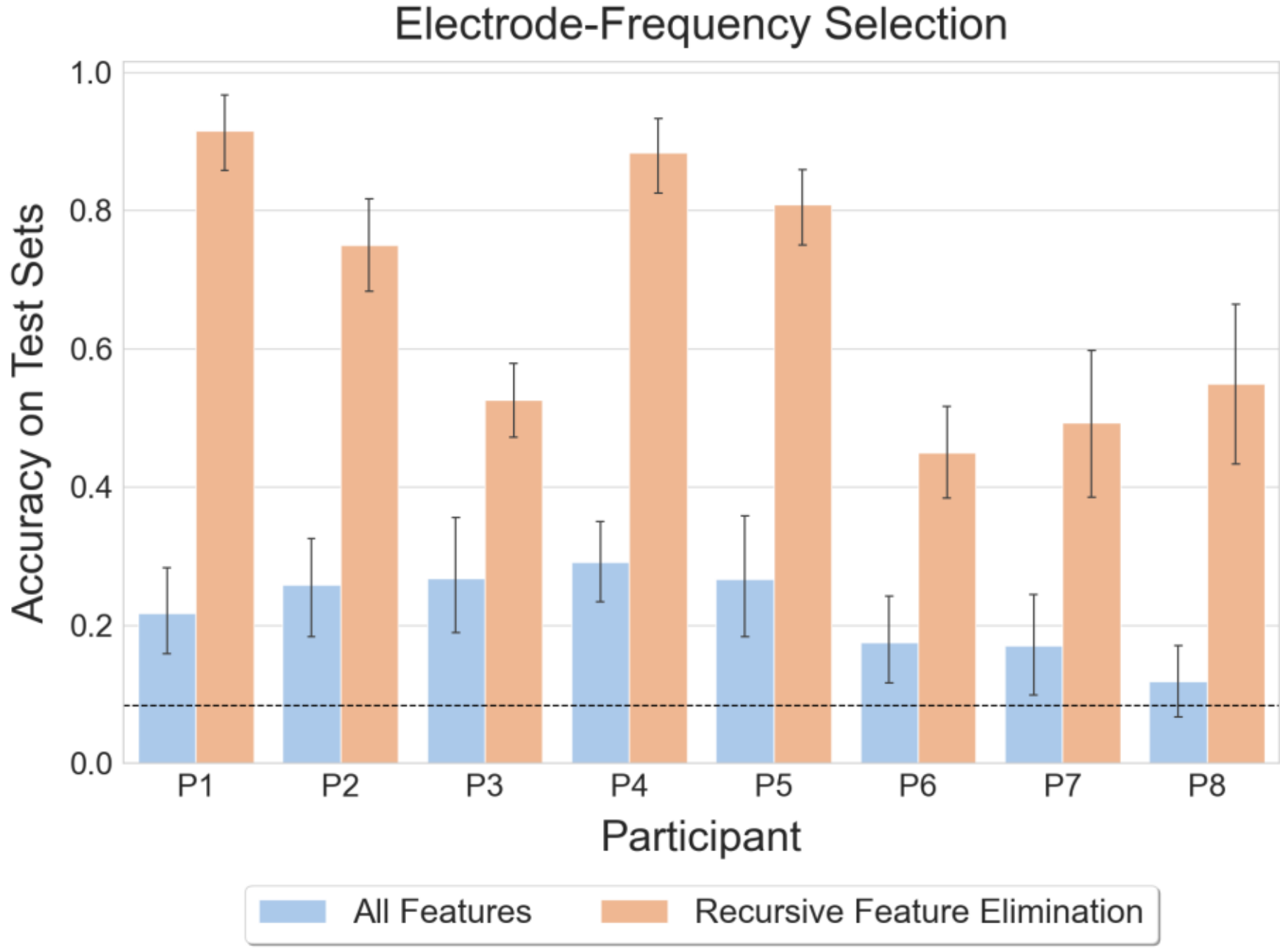

**Figure 2.** Test-set classification accuracies for combined electrode-frequency selection using recursive feature elimination (RFE). For each participant, blue bars show the mean cross-validated accuracy without feature selection, while orange bars indicate performance after applying RFE. Error bars represent the standard deviation across cross-validation folds.

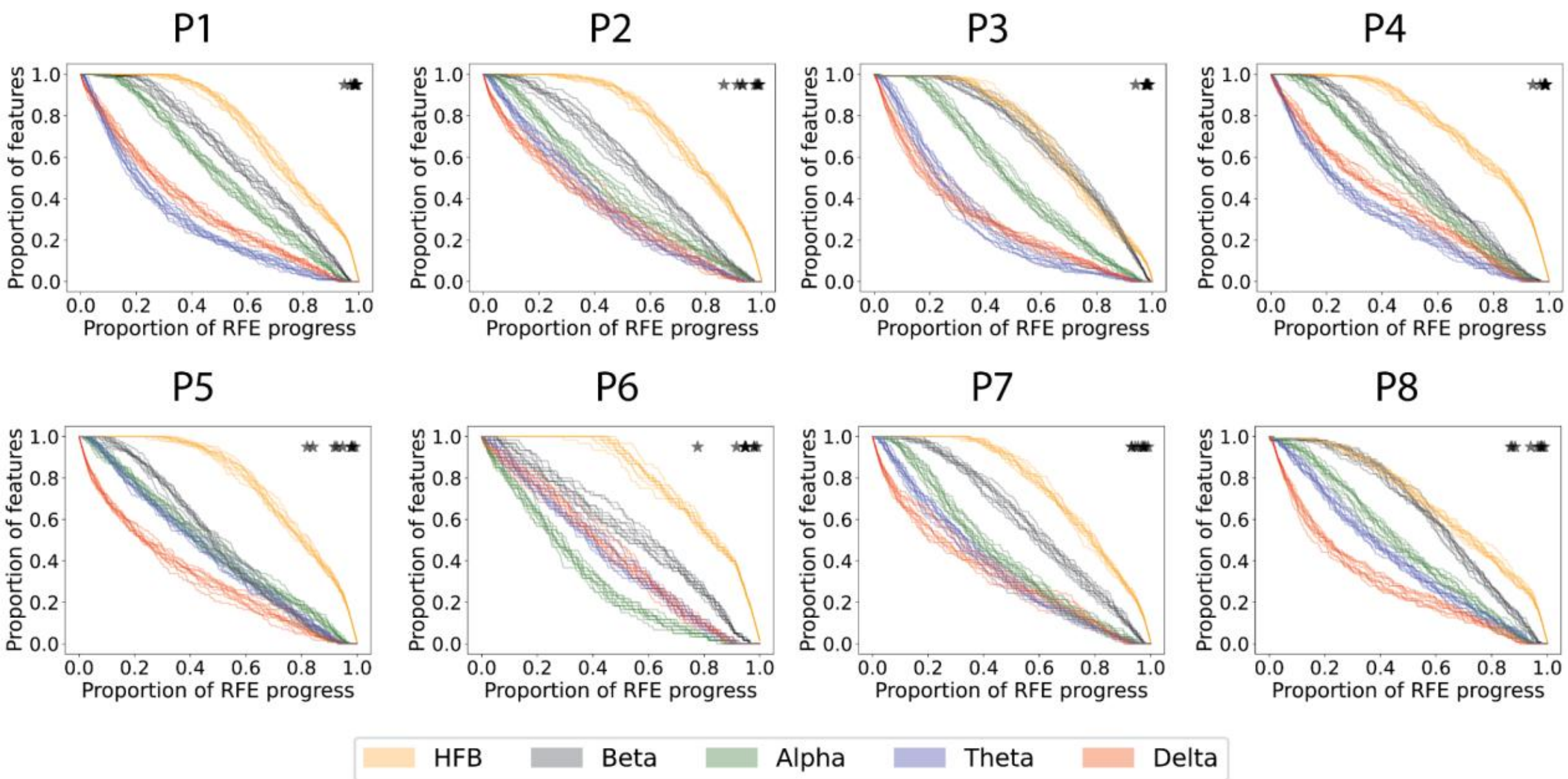

**Figure 3.** Elimination of frequency band features during combined electrode-frequency selection per participant using recursive feature elimination (RFE). The stars, shown at the top of each plot, indicate the point along the x-axis (RFE progression) at which classification performance was highest for each cross-validation fold. The later in the process a feature gets eliminated, the more important it is for the classification. On average, HFB electrodes were eliminated last, indicating the contribution of the most relevant information for classification.

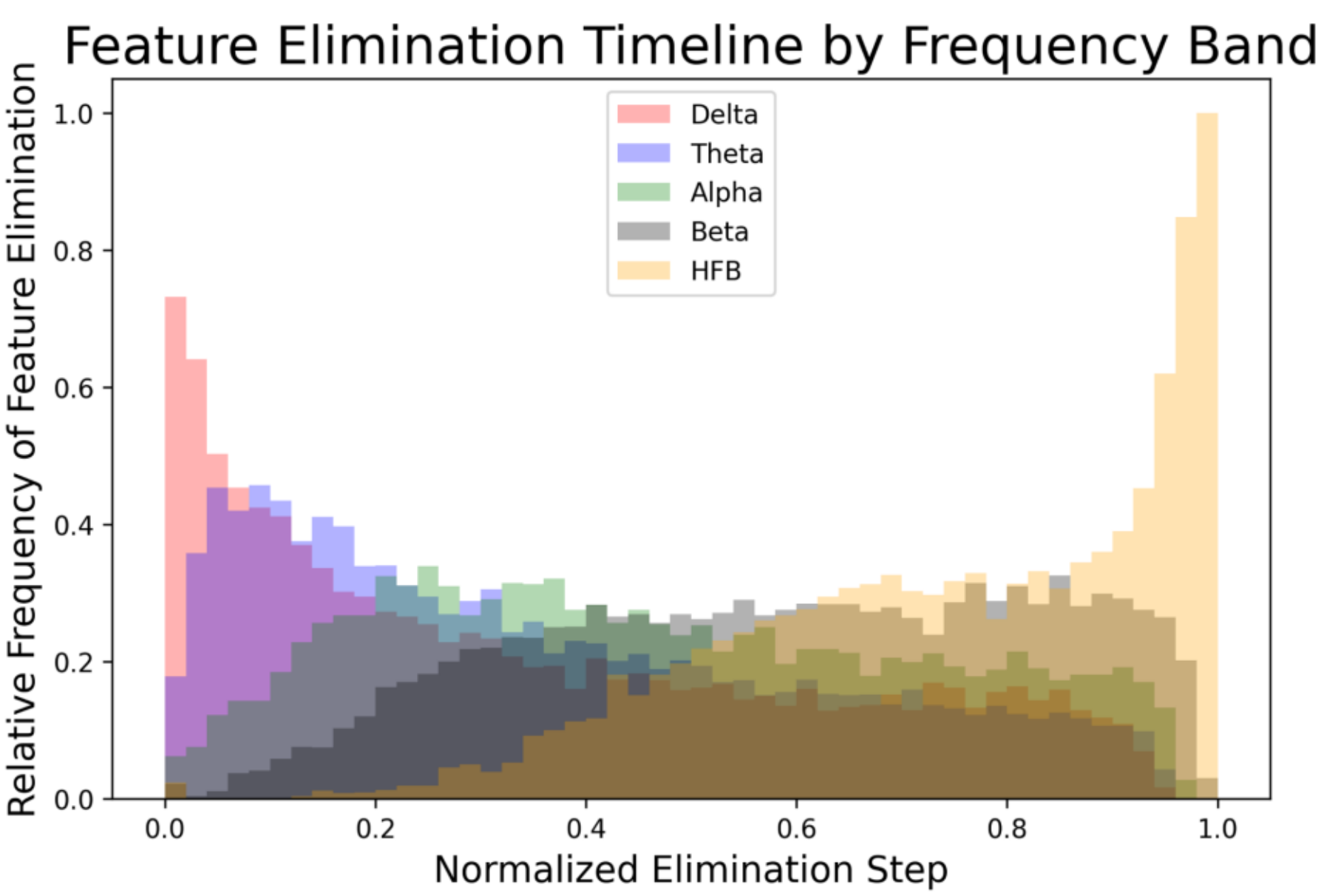

**Figure 4.** Elimination per frequency band in combined electrode-frequency selection using recursive feature elimination across participants. The proportion of each frequency band removed at each normalized elimination step is shown. HFB is eliminated last, indicating that it carries the most informative features for classification.

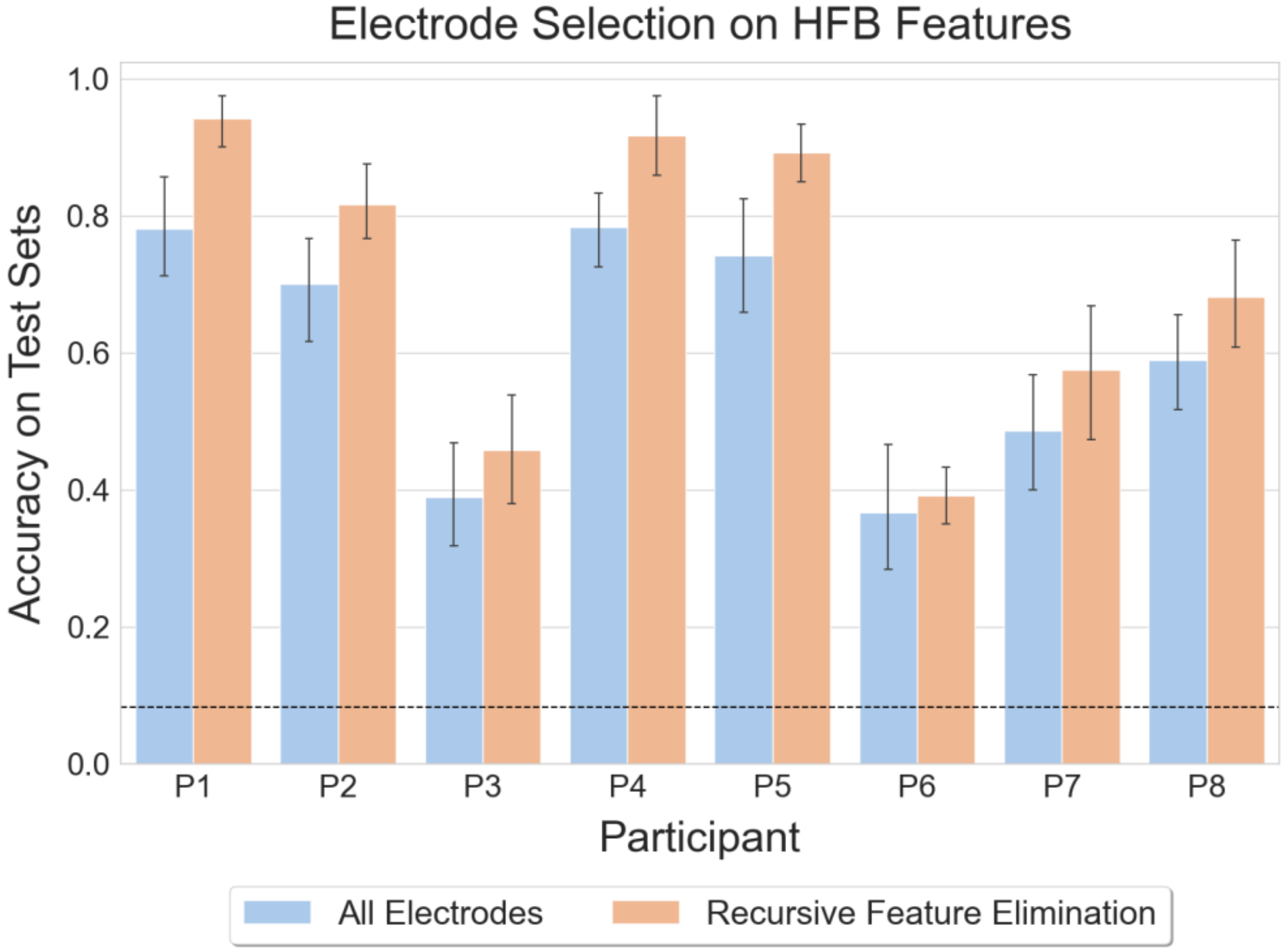

**Figure 5.** Test-set accuracies for electrode selection using recursive feature elimination (RFE). For each participant, blue bars show the mean cross-validated accuracy on the HFB features without electrode selection, while orange bars indicate performance after applying RFE on the electrode dimension. Error bars represent the standard deviation across cross-validation folds.

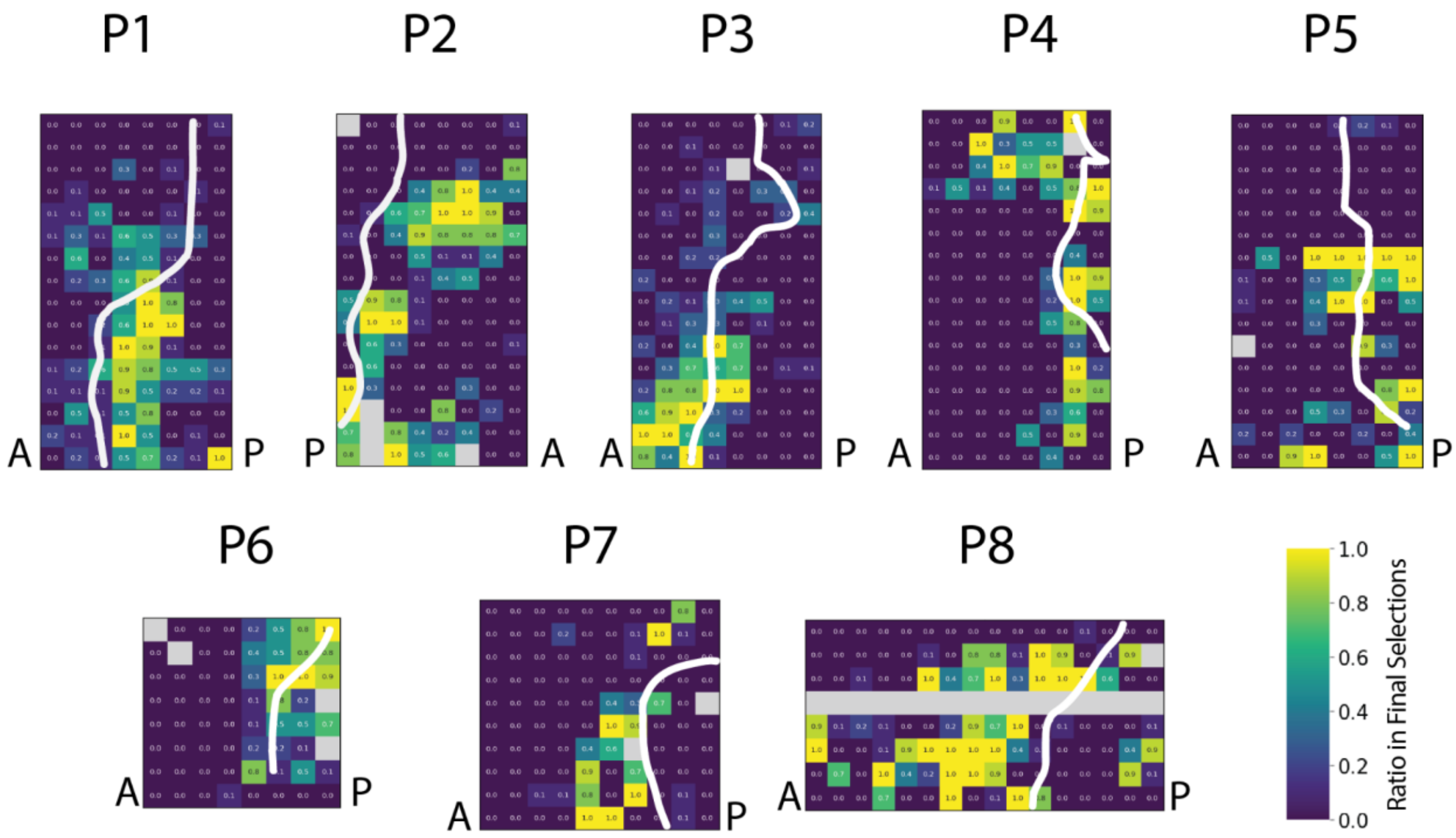

**Figure 6.** Electrodes selected by recursive feature elimination for each participant. The color scale indicates the ratio of cross-validation folds (out of 10) in which a given electrode appeared in the final feature selection. electrodes excluded from analysis due to noisiness or flat signals are shown in gray, and the central sulci are outlined in white. “A” and “P” denote anterior and posterior directions on the cortex of participants.

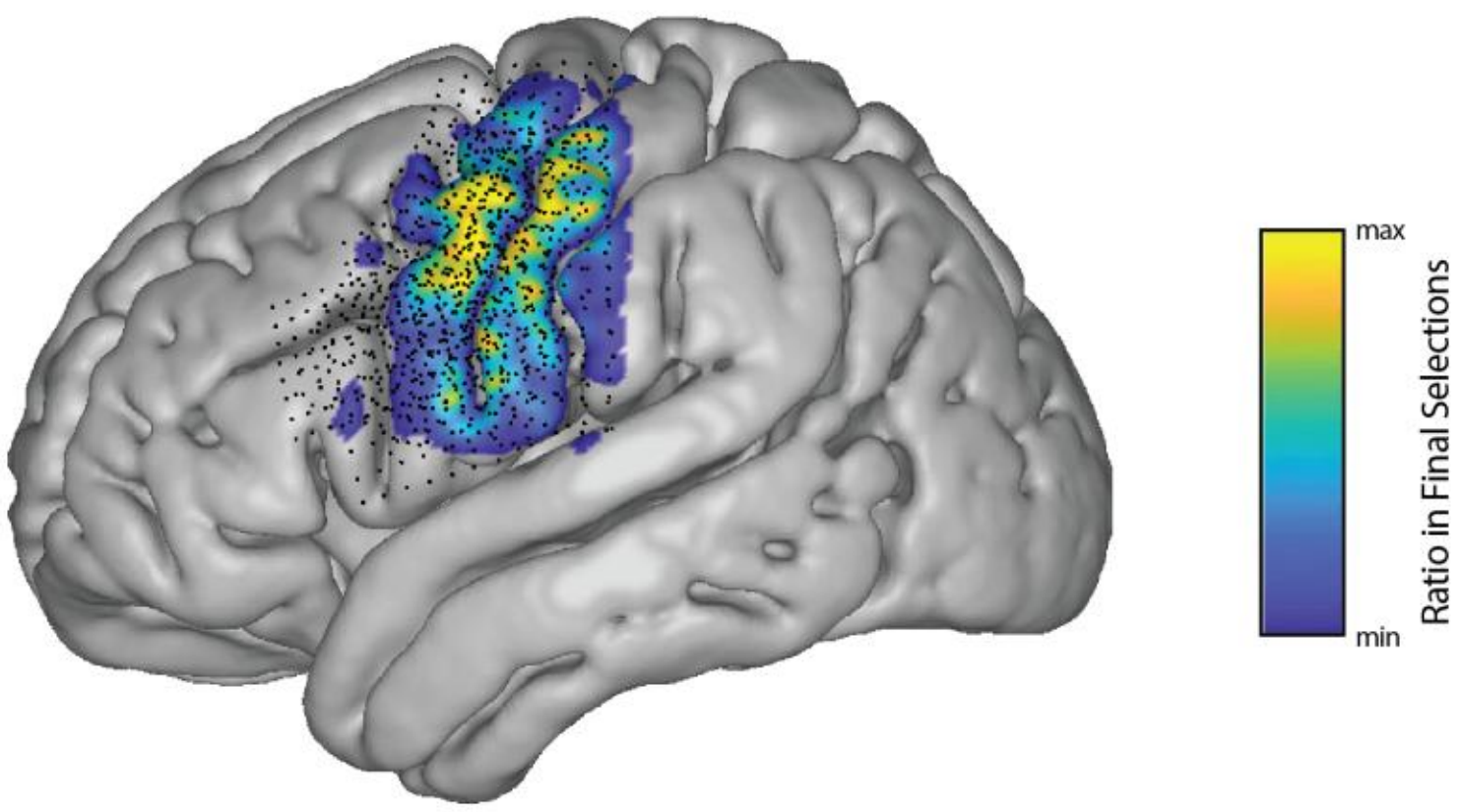

**Figure 7.** Electrodes selected by recursive feature elimination for all participants, projected onto a common MNI brain. The color scale indicates the cumulative ratio of cross-validation folds (out of 10) in which a given electrode appeared in the final feature selection across participants. For visualization purposes, data from P2 is mapped onto the left hemisphere.

Finally, RFE was applied solely to the time dimension, using HFB features derived from all non-noisy electrodes. A temporal window from 1 second before to 1 second after the voice onset of each trial was used, sampled at 50 Hz, resulting in 100 candidate time points that were iteratively removed. RFE led to a significant improvement in classification accuracy as compared to models using all available time points (from 66±13% to 71±11%, $Z_{all\text{-}selected}$= -2.5, p = 0.01, Wilcoxon signed-rank test, Figure 8, Supplementary Table 1 Columns 3 & 5). The outcomes for individual participants, as well as the pooled median across participants, are shown in Figure 9. Across participants, the most consistently selected time points clustered shortly after voice onset, with a weighted average of 330±170 ms. In addition, some informative time points were selected immediately before voice onset, with the earliest selected points occurring approximately 150 ± 120 ms prior to voice onset. These near-onset pre-speech signals likely reflect preparatory activity relevant for word discrimination.

In contrast, selections occurring much earlier in the trial, i.e. more than 0.5s before voice onset, were generally associated with lower decoding performance and likely reflect noise or task-unrelated activity. Participants with reduced accuracies tended to show a larger proportion of such early selections. For example, P3, P6, and P8 achieved mean respective accuracies of 36%, 39%, and 65%, and in these participants 40%, 24%, and 24% of time points between 1 and 0.5 s before voice onset were selected, respectively. By comparison, P1, P4, and P5 achieved the highest mean accuracies (87%, 91%, and 88%) and had few or no selections in this interval (0%, 0% and 4%).

At the group level, the median pattern revealed moderate selection of the time points right before voice onset, followed by a strong concentration of selected time points immediately after onset, and a decline toward the end of the window, when articulation was completed and provided less discriminative information (Figure 9B).

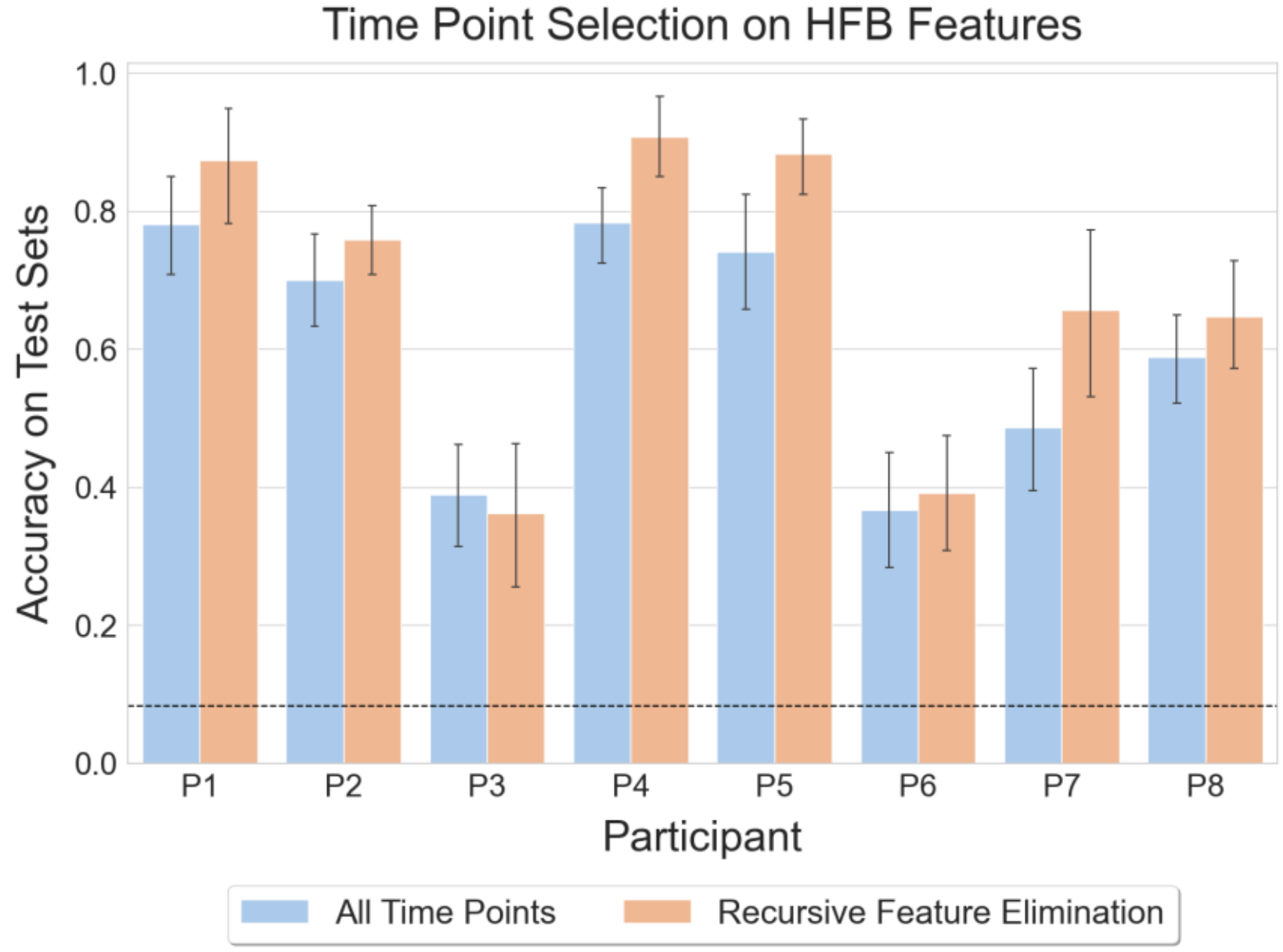

**Figure 8.** Test-set accuracies for time point selection using recursive feature elimination (RFE). For each participant, blue bars show the mean cross-validated accuracy with all time points, while orange bars indicate performance after applying RFE on the time dimension. Error bars represent the standard deviation across cross-validation folds.

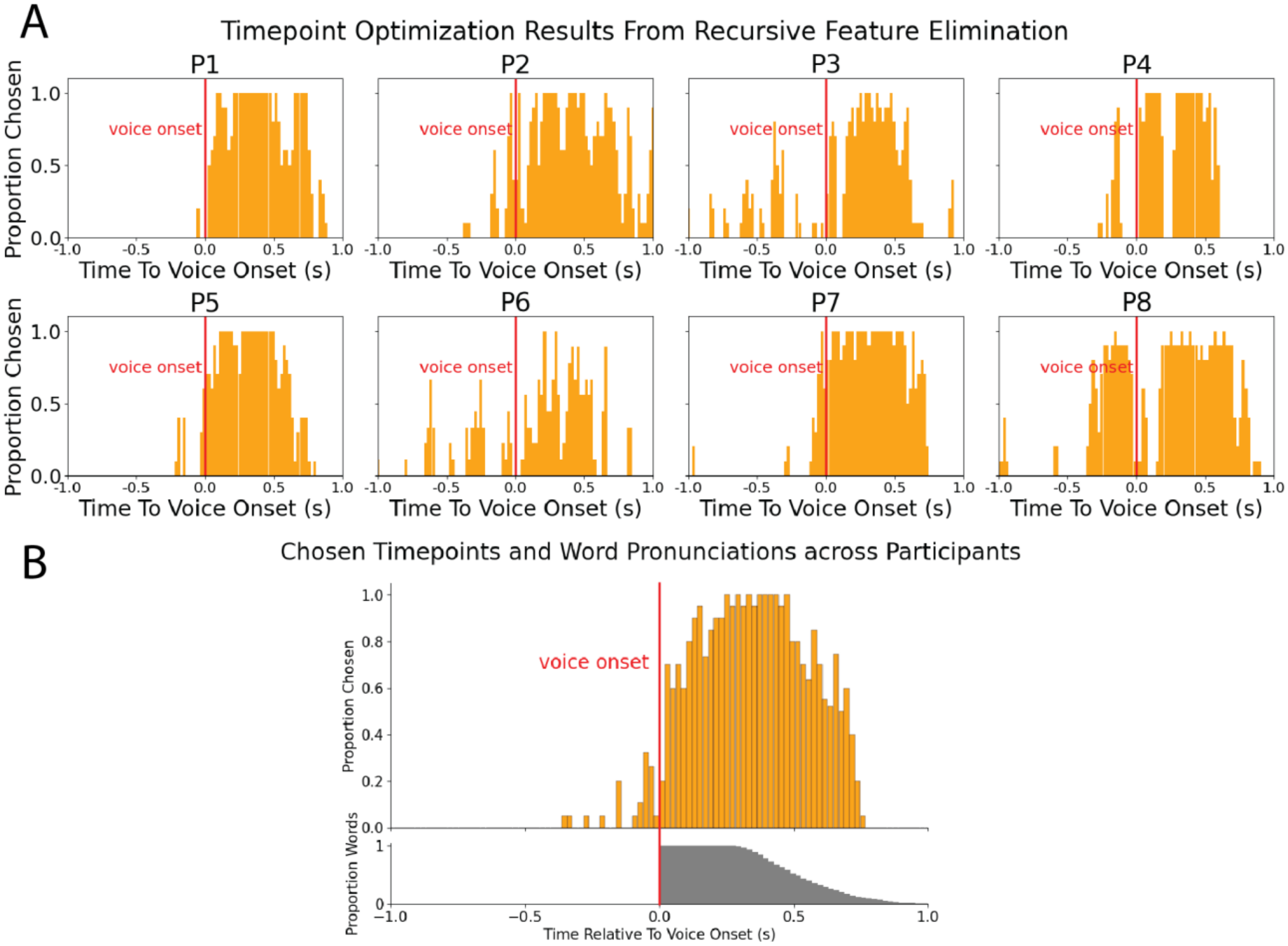

**Figure 9. Time points selected by recursive feature elimination.** A) Proportion of cross-validation folds (out of 10) in which each time point was retained in the final feature selection, shown separately for each participant. The red line indicates voice onset per trial. B) Top: Median proportion of selected time points, pooled across all participants. Bottom: Distribution of spoken words across time, showing the proportion of trials across participants with vocalization at each time point.

*3.2 Retrieving relevant information in a limited task setup*

In many BCI studies, prior knowledge about neural signals can guide manual feature selection. For example, HFB activity is known to correlate strongly with motor and speech processes, and voice onset times can be used to temporally align neural activity to speech production. However, such external information is not always available to guide feature selection. For example, in communication BCIs for individuals with locked-in syndrome, voice onset times are typically unavailable, which can reduce decoding performance (Rabbani et al., 2025). As a next step, we evaluated whether BCI-sift could determine the optimal feature selection and possibly mitigate performance loss under such conditions. For this, we repeated the time dimension selection without incorporating information about voice onset times. A window from 0.5 seconds before to 2 seconds after the cue was used for each trial. Participant P3 was excluded from this analysis due to missing cue signals in the recorded data. Classification accuracy again improved significantly with feature selection compared to the inclusion of all time points (from 50±8% to 58±13%, $Z_{all\text{-}selected}$= -2.4, p = 0.02, Wilcoxon signed-rank test, Supplementary Table 1 Columns 6 & 7). In addition, the performance after RFE was not significantly lower than in the onset-aligned analysis (from 67±25% to 58±13%, $Z_{onset\text{-}cue}$= -1.4, p = 0.18, Wilcoxon signed-rank test, Supplementary Table 1 Columns 5 (excl. P3) & 7). The selected time points began slightly before the voice onset times, likely reflecting preparatory information, and coincided with the periods of word production, as in the voice-onset based analysis (Figure 10). Participant P7 exhibited the highest variability in selected time points, which corresponded with a large variance in cue-to-onset latency in this participant.

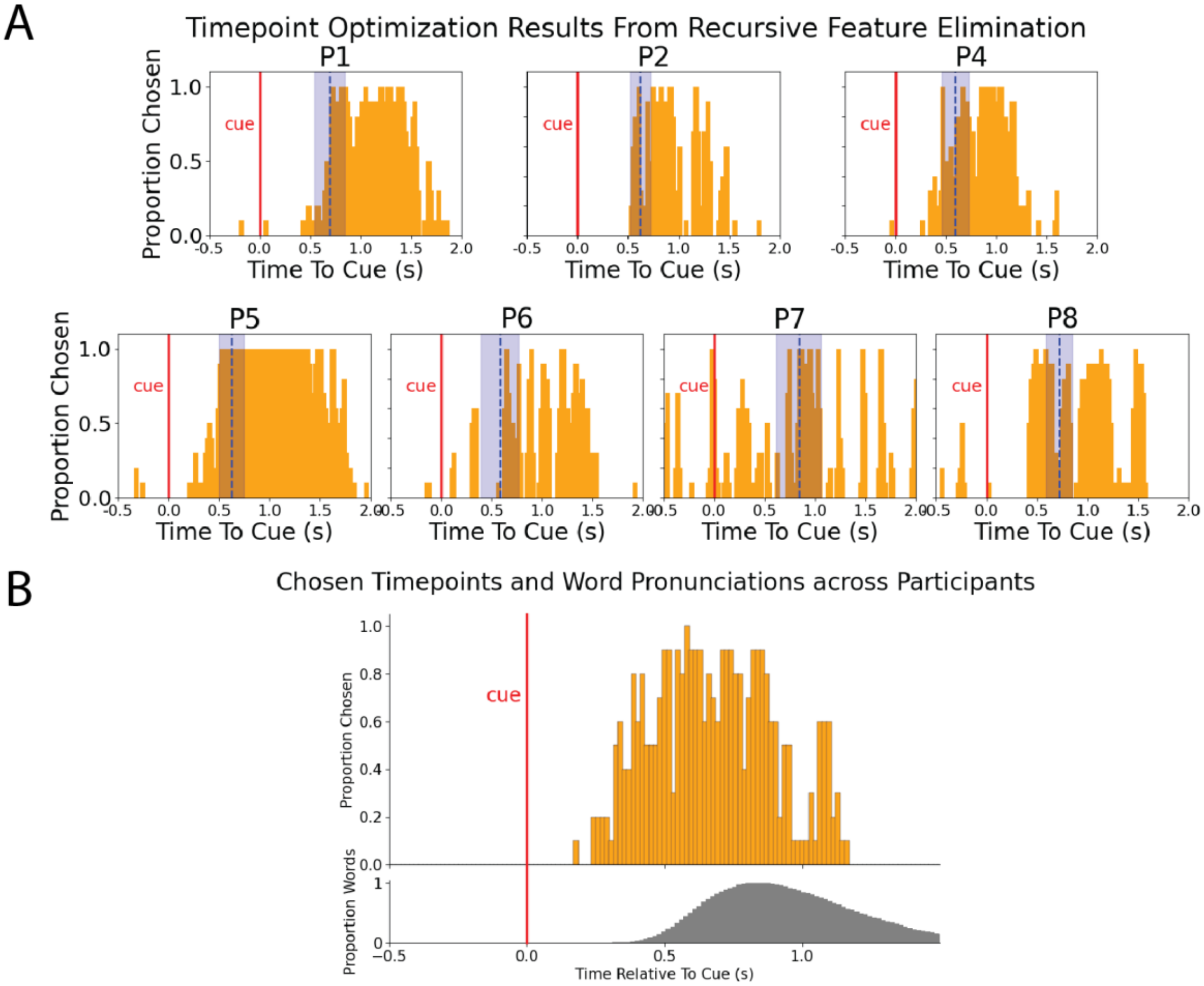

**Figure 10. Time points selected by recursive feature elimination on cue-aligned data.** A) Proportion of cross-validation folds (out of 10) in which each time point was retained in the final feature selection, shown separately for each participant. The red line indicates the cue per trial. The dashed blue line indicates the mean voice onset time, and the shaded blue region is the standard deviation of voice onset times across trials. P3 was excluded from this analysis due to missing cue timing information. B) Top: Median proportion of selected time points, pooled across all participants. Bottom: Distribution of spoken words across time, showing the proportion of trials with vocalization at each time point.

*3.3 Automatic data denoising*

Beyond optimizing feature selection when external task information is unavailable, BCI-sift can also help address data quality issues. A possible use-case for BCI-sift is the automatic identification and exclusion of noisy or flat electrodes. In the subsequent analysis, electrodes that had previously been excluded due to excessive noise or having flat signals were reintroduced, and the electrode dimension was optimized again. Participant P1, who had no electrodes labeled as noisy or flat, was excluded from this analysis. Across the remaining seven participants, a total of 32 electrodes had been marked as noisy. Recursive feature elimination again yielded a significant improvement in accuracy compared to the inclusion of all electrodes (from 45±20% to 61±28%, $Z_{all\text{-}selected}$= -2.2, p = 0.03, Wilcoxon signed-rank test, Figure 11, Supplementary Table 1 Columns 8 & 9). The resulting accuracy did not differ significantly from the analysis in which noisy electrodes were excluded a priori (from 67±21% to 61±28%, $Z_{exluded\text{-}included}$= -1.8, p = 0.08, Wilcoxon signed-rank test, Figure 11, Supplementary Table 1 Columns 4 (excl. P1) & 9).The only participants in whom manual exclusion of noisy electrodes outperformed recursive feature elimination were those with the highest proportions of noisy electrodes relative to the total number of electrodes (6.3% for P6 and 13.3% for P8). On average, the 32 noisy electrodes were included in the final selections in only 4.1% of cases, compared to 22.6% for the remaining electrodes. Notably, 30 of the 32 noisy electrodes were not retained in any of the final feature selections across all cross-validation folds.

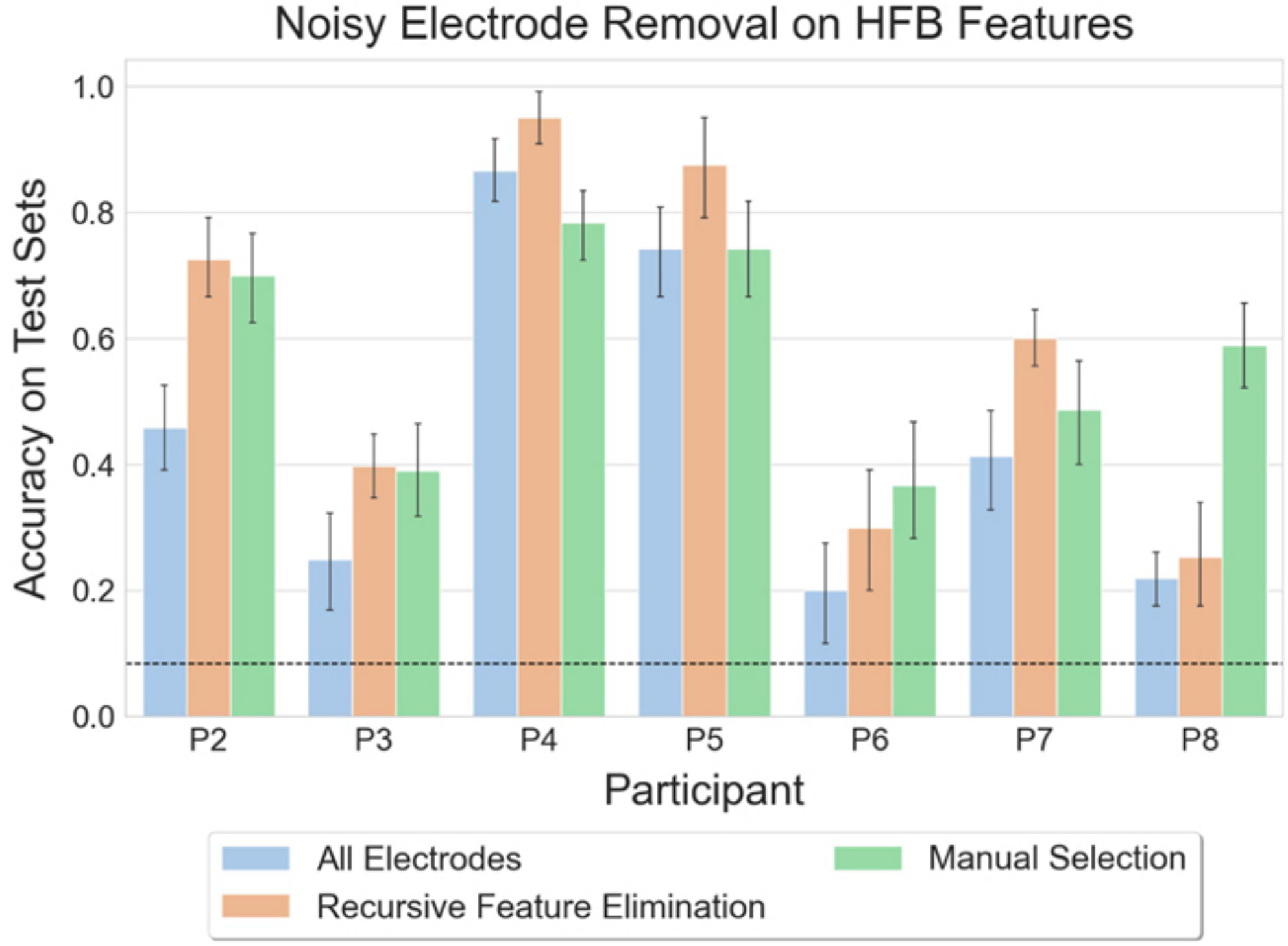

**Figure 11.** Test-set accuracies for electrode selection analysis including electrodes previously marked as noisy. For each participant, blue bars indicate the mean cross-validated accuracy without feature selection, orange bars show performance after applying recursive feature elimination, and green bars show performance when noisy electrodes were excluded manually. Participant P1 was excluded because no electrodes were labeled as noisy.

### *3.4 Comparison of optimization algorithms in BCI-sift*

Although most of the results have highlighted RFE as a representative example, the BCI-sift toolbox supports multiple optimization strategies for feature selection. When applied to the electrode dimension, as in Section 3.1, these alternative algorithms produced broadly comparable outcomes (Figure 12). The average Pearson correlation between the selection outcomes of any two methods was r=0.42, while the mean correlation between algorithm outputs and electrode density was negligible (r = 0.006). In this particular dataset, RFE achieved the most effective feature selection and yielded the highest decoding performance. However, for other datasets or signal characteristics, alternative optimization approaches included in BCI-sift may provide superior results.

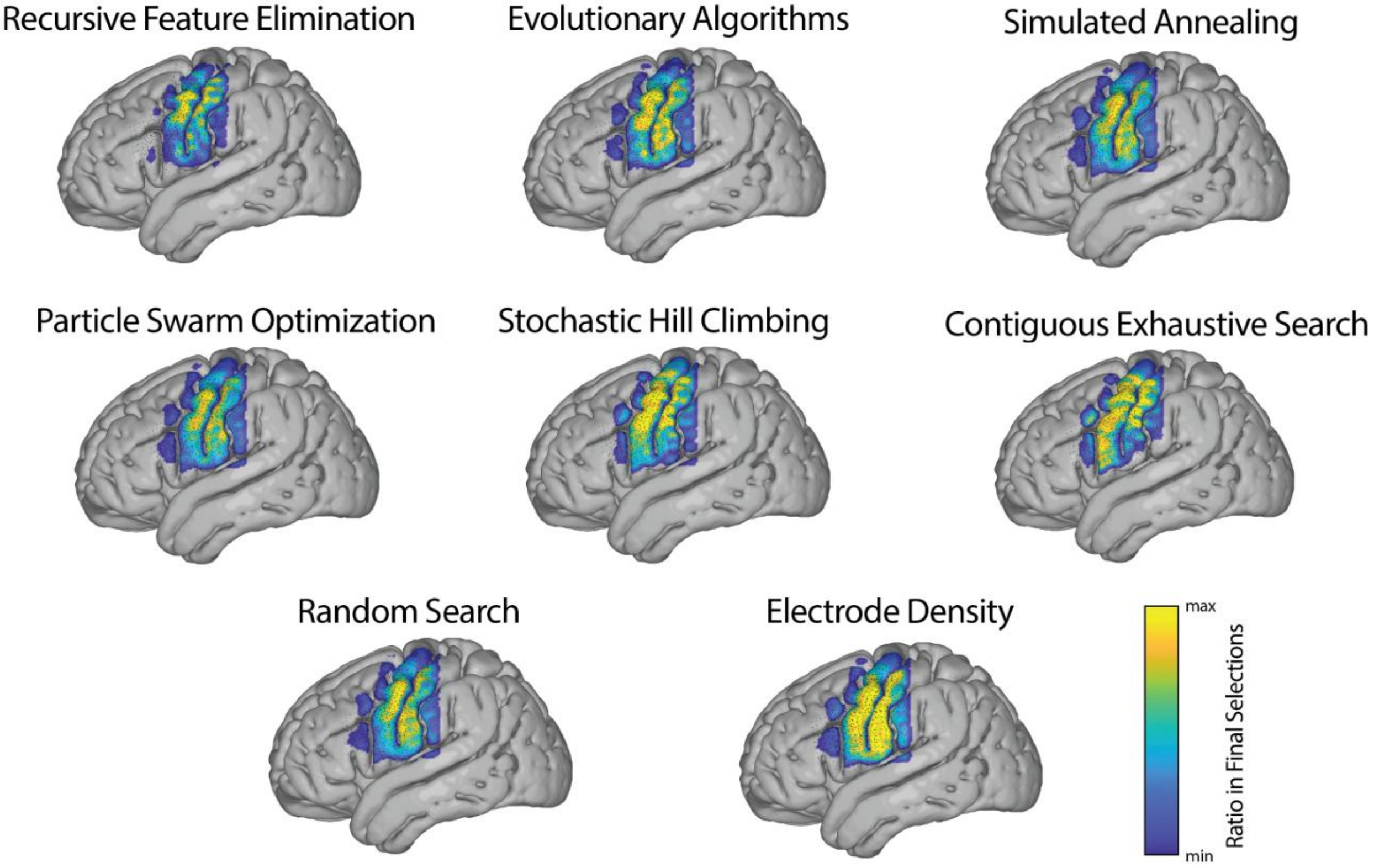

**Figure 12.** Results of electrode feature selection, showing electrodes selected by each optimization algorithm for channel selection across all participants, projected onto a common MNI brain. The color scale indicates the number of cross-validation folds (out of 10) in which a given electrode appeared in the final feature set; for the electrode-density visualization, each electrode was instead assigned a fixed value of 5. For visualization purposes, data from P2 is mapped onto the left hemisphere.

**4. Discussion**

In this work, we introduced BCI-sift – a modular, Python-based toolbox designed to streamline systematic and interpretable feature selection for BCI applications. By integrating a range of established optimization algorithms BCI-sift provides a flexible and extensible framework for identifying the most informative features across multiple dimensions such as electrode, time, and signal frequency. This results in improved decoding accuracy, improved feature interpretability, the ability to retrieve information in situations with limited task knowledge and enabling automatic data denoising.

Our analyses using HD ECoG data from a word pronunciation task show that optimization reliably enhances decoding performance across participants and across all feature dimensions examined. In addition to improving accuracy, the toolbox provides interpretability by identifying the most informative features. Specifically, RFE emphasized the relevance of high frequency band activity near the central sulcus and within peri-speech time intervals, aligning with prior findings about neural correlates of speech production. The ability of the toolbox to recover such physiologically meaningful patterns highlights the interpretability of its outputs and their value for both hypothesis-driven neuroscience research and BCI development. Furthermore, the validation demonstrated that the toolbox could recover information related to behavioral timing, such as voice onset, and automatically detect and exclude noisy electrodes, offering practical benefits in scenarios where prior knowledge is limited.

Another strength of BCI-sift lies in its modular and scikit-learn–compatible design. This ensures that it can be easily incorporated into existing workflows, extended with novel optimization or preprocessing methods, and applied to a wide range of classifiers or other prediction models, including more advanced algorithms such as non-linear neural networks or clustering algorithms. Furthermore, the inclusion of hyperparameter tuning and parallelized processing enhances usability for both exploratory research and large-scale applications. The toolbox not only produces quantitative performance metrics but also automatically generates visualizations, facilitating transparent and intuitive interpretation of results.

In this work, we demonstrated results on a single dataset and primarily with one algorithm, recursive feature elimination. This choice was deliberate: our aim was not to compare algorithms against one another on a specific dataset, but rather to introduce a flexible framework that integrates a broad range of optimization approaches and to illustrate its functionality in principle. Results from another algorithm, contiguous exhaustive search, have been reported separately (Offenberg et al., 2025), and future work will extend the evaluation of the full algorithm set across additional datasets and modalities. It should also be noted that the magnitude of accuracy improvements may depend on factors such as data preprocessing and preparation.

BCI-sift provides a range of additional functionalities left out of the scope of this report. Owing to its compatibility with scikit-learn, it can be seamlessly integrated into existing workflows and used with any prediction model supported by that framework, offering substantial flexibility. It supports parallelized processing across multiple CPU cores, enabling efficient use in both cloud-based and high-performance computing environments. An integrated hyperparameter tuning module allows automated optimization of algorithm-specific parameters, such as the temperature schedule in simulated annealing or the regularization strength in SVMs. The modular design further facilitates the addition of custom algorithms or preprocessing steps, and the project is intended to be community-driven, encouraging external contributions. The toolbox also generates informative outputs, including plots and tables that support intuitive result interpretation. While validation here was performed with HD ECoG recordings, the same framework is directly applicable to other modalities, including implanted and non-implanted BCI data.

In summary, BCI-sift addresses the fragmented landscape of feature selection in BCI research by providing a unified, extensible, and interpretable Python toolbox. Our validation experiments confirm that systematic feature optimization not only improves classification accuracy but also highlights physiologically meaningful patterns. We anticipate that BCI-sift will support both methodological advances and practical applications in neuroscience and BCI development, while growing as a community resource for feature selection and optimization.

Supplementary Table 1: Decoding accuracies per participant across conditions.

| | Clean, All Freq No FS VO | Clean, All Freq FS: E+F VO | Clean, HFB No FS VO | Clean, HFB FS: E VO | Clean, HFB FS: T VO | Clean, HFB No FS Cue | Clean, HFB FS: T Cue | All, HFB No FS VO | All, HFB FS: E VO |
|---|---|---|---|---|---|---|---|---|---|
| P1 | $17 \pm 8\%$ | $92 \pm 8\%$ | $79 \pm 13\%$ | $96 \pm 4\%$ | $92 \pm 8\%$ | $50 \pm 8\%$ | $66 \pm 7\%$ | — | — |
| P2 | $29 \pm 4\%$ | $75 \pm 8\%$ | $71 \pm 13\%$ | $83 \pm 8\%$ | $75 \pm 4\%$ | $50 \pm 8\%$ | $58 \pm 8\%$ | $46 \pm 8\%$ | $71 \pm 4\%$ |
| P3 | $22 \pm 8\%$ | $52 \pm 4\%$ | $36 \pm 11\%$ | $46 \pm 10\%$ | $39 \pm 13\%$ | — | — | $25 \pm 6\%$ | $39 \pm 8\%$ |
| P4 | $29 \pm 13\%$ | $88 \pm 8\%$ | $79 \pm 4\%$ | $96 \pm 4\%$ | $92 \pm 8\%$ | $58 \pm 8\%$ | $71 \pm 8\%$ | $92 \pm 8\%$ | $96 \pm 4\%$ |
| P5 | $25 \pm 13\%$ | $83 \pm 8\%$ | $75 \pm 8\%$ | $92 \pm 8\%$ | $92 \pm 8\%$ | $58 \pm 4\%$ | $75 \pm 8\%$ | $75 \pm 8\%$ | $92 \pm 8\%$ |
| P6 | $17 \pm 8\%$ | $42 \pm 8\%$ | $33 \pm 8\%$ | $42 \pm 8\%$ | $42 \pm 8\%$ | $22 \pm 8\%$ | $26 \pm 8\%$ | $17 \pm 4\%$ | $33 \pm 13\%$ |
| P7 | $17 \pm 9\%$ | $41 \pm 11\%$ | $50 \pm 5\%$ | $64 \pm 9\%$ | $66 \pm 18\%$ | $22 \pm 8\%$ | $32 \pm 11\%$ | $45 \pm 6\%$ | $61 \pm 7\%$ |
| P8 | $8 \pm 8\%$ | $58 \pm 8\%$ | $61 \pm 6\%$ | $67 \pm 8\%$ | $63 \pm 11\%$ | $33 \pm 14\%$ | $50 \pm 17\%$ | $25 \pm 4\%$ | $21 \pm 5\%$ |
| **Median** | $\mathbf{19 \pm 4\%}$ | $\mathbf{67 \pm 19\%}$ | $\mathbf{66 \pm 13\%}$ | $\mathbf{75 \pm 19\%}$ | $\mathbf{71 \pm 11\%}$ | $\mathbf{50 \pm 8\%}$ | $\mathbf{58 \pm 13\%}$ | $\mathbf{45 \pm 20\%}$ | $\mathbf{61 \pm 28\%}$ |

VO = voice onset aligned; Cue = cue aligned; Clean = noisy electrodes manually removed; All = all electrodes included; FS = feature selection; E = electrodes; F = frequency; T = time.